\documentclass[a4paper]{jpconf}
\usepackage{pslatex}
\usepackage[T1]{fontenc}
\usepackage{graphicx}
\usepackage{epsfig}

\usepackage{calc}
\usepackage{ifthen}
\usepackage{subfigure}

\newcommand{\bea}{\begin{eqnarray}}
\newcommand{\eea}{\end{eqnarray}}

\newcommand{\nc}{\newcommand}
\nc{\renc}{\renewcommand}
\nc{\eqs}[2]{\mbox{Eqs.~(\ref{#1},\,\ref{#2})}}
\nc{\eq}[1]{\mbox{Eq.~(\ref{#1})}}
\nc{\figs}[2]{\mbox{Figs.~(\ref{#1},\,\ref{#2})}}
\nc{\fig}[1]{\mbox{Fig~.(\ref{#1})}}
\nc{\be}[1]{\begin{equation} \mbox{$\label{#1}$}}
\nc{\ee}{\vspace{0.1cm}\end{equation}}

\newcommand{\bean}{\begin{eqnarray*}}
\newcommand{\eean}{\end{eqnarray*}}

\def\GeV{{\rm \ GeV}}

\def\TeV{{\rm \ TeV}}

\def\lae{\;^{<}_{\sim} \;} \def\gae{\; ^{>}_{\sim} \;}

\def\pb{\phi_{B}}

\def\pbh{\hat{\phi}_{B}}

\def\sp{\hat{s}}

\def\vs1{\vspace{0.1cm}}

\nc{\npp}[3]{{\it  Nucl.\ Phys.\ }{{\bf #1} {(#2)} {#3}}}
\nc{\prdd}[3]{{\it  Phys.\ Rev.\ D\ }{{\bf #1} {(#2)} {#3}}}
\nc{\prll}[3]{{\it Phys.\ Rev.\ Lett.\ }{{\bf #1} {(#2)} {#3}}}
\nc{\pll}[3]{{\it  Phys.\ Lett.\ }{{\bf #1} {(#2)} {#3}}}

\begin{document}
\title{Generation of WIMP Miracle-like Densities of Baryons and Dark Matter
}
\author{John McDonald}

\address{Consortium for Fundamental Physics, Cosmology and Astroparticle Physics Group, Department of Physics, University of
Lancaster, Lancaster LA1 4YB, UK}

\ead{j.mcdonald@lancaster.ac.uk}

\begin{abstract}

      The observed density of dark matter is of the magnitude expected for a thermal relic weakly-interacting massive particle (WIMP). In addition, the observed baryon density is within an order of magnitude of the dark matter density. This suggests that the baryon density is physically related to a typical thermal relic WIMP dark matter density. We present a model which simultaneously generates thermal relic WIMP-like densities for both baryons and dark matter by modifying a large initial baryon asymmetry. Production of unstable scalars carrying baryon number at the LHC would be a clear signature of the model.  

\vs1
\noindent (Talk presented at DSU 2011, KITPC, Beijing, China.) 

 \end{abstract}


     The ratio of the mass density of baryons to that of dark matter (the BDM ratio) is observed to be $\Omega_{B}/\Omega_{DM} \approx 1/5$. However, in most models the physics of baryogenesis and of dark matter production is physically unrelated. Why then is the density
in baryons within an order of magnitude of that in dark matter? Either: 

\vspace{0.1cm} 
\noindent $\bullet$ A remarkable coincidence

\vspace{0.1cm} 
\noindent $\bullet$ An anthropic selection mechanism

\vspace{0.1cm} 
\noindent $\bullet$ The physics of the observed baryon asymmetry and dark matter densities is in some way related.

\vspace{0.1cm} 
\noindent Therefore the BDM ratio may be a powerful clue to the correct particle physics theory. 

\vs1
Up to now there have been broadly two classes of model:

\vspace{0.1cm} 
\noindent 1). Direct mechanism: The dark matter particle and baryon number densities are directly related by a conserved charge, 
$$ \Rightarrow Q_{tot} = Q_{B} + Q_{X} \Rightarrow n_{cdm} \sim n_{B} \Rightarrow M_{DM} \sim m_{n} n_{B}/n_{cdm} \sim 1-10 \GeV$$ 
There are many models in this catagory, which typically predict 
asymmetric dark matter \cite{direct0,direct1,direct2,direct3,direct4,adm,pangenesis,pangenesis2}. 
Models also exist which break the simple mass relation $M_{DM} \sim 1-10 \GeV$, see \cite{vdirect1,vdirect2,vdirect3,vdirect4,vdirect5,vdirect6}\footnote{In \cite{assim} a model is discussed where the correct dark matter density is generated via its interaction with the baryon asymmetry.}.

\vspace{0.1cm} 
\noindent 2). Indirect mechanism: The dark matter and baryon densities are related by {\it similar} but {\it seperate} physical mechanisms for their origin $\Rightarrow$ Less rigid relation between $n_{B}$ and $n_{cdm}$; can have $n_{B} \gg n_{cdm} \Rightarrow$ larger $M_{DM}$. An example based on $d=4$ Affleck-Dine leptogenesis is given in \cite{rhsn}, where dark matter is due to a condensate of right-handed sneutrinos \footnote{The author is not aware of other examples in this class and would be interested to know if other models exist.}

\vs1
\noindent However, there are in fact {\it two} seperate coincidences. 
\vs1
\noindent 1). Why are the baryon and dark matter densities similar to each other? 
\vs1
\noindent 2). Why are they \underline{both} similar to the "WIMP Miracle" density? (Most models do not address 2.)

       The "WIMP Miracle" refers to the natural similarity of the observed dark matter density
and the thermal relic density of particles with weak-scale masses and 
interactions. This is widely 
interpreted as a strong indication that dark matter is due to thermal relic WIMPs. For dark matter with constant $< \sigma v_{rel}>$, where $\sigma$ is the non-relativistic annihilation cross-section, the observed dark matter density requires 
$$ < \sigma v_{rel}> \approx 10^{-9} \GeV^{-2}$$ 
The DM annihilation cross-section may be expresed in terms of the annihilation matrix element ${\cal M}$ as 
\be{e1}  <\sigma v_{rel}> = \frac{|{\cal M}|^2}{32 \pi m_{DM}^2}    ~.\ee
The matrix element is dimensionless and typically has the form 
$|{\cal M}|^{2} = g_{eff}^4$. where $g_{eff}$ is an effective coupling constant. Therefore 
 \be{e2}  <\sigma v_{rel}> = 1.6 \times 10^{-9} 
\left(\frac{g_{eff}}{0.2}\right)^{4}  
\left(\frac{100 \GeV}{m_{DM}}\right)^{2}
\GeV^{-2} 
~.\ee
Therefore if $M_{DM}$ is in the range $100 \GeV - 1 \TeV$ and if $g_{eff}$ is not too much smaller than the weak interaction coupling $g \approx 0.65$ (which is natural as $g_{eff}$ can have additional factors from mixing angles, mass ratios etc), then we get the right amount of thermal relic dark matter. 

   If the WIMP miracle does explain dark matter, and if we discount coincidence and antropic selection, then we need to explain why the baryon asymmetry is also similar to the WIMP miracle density. The similarity to the dark matter density then {\it follows} from this 
i.e. the observed baryon to dark matter ratio is actually due to a more fundamental relation between each of the 
densities and the WIMP miracle density.    

    In \cite{bm1} and \cite{bm2} we presented one approach to making this connection, based on a mechanism we call {\it baryomorphosis}. [Recently an alternative approach to achieving a WIMP-like baryon asymmetry, {\it WIMPy baryogenesis}, has been proposed in \cite{wimpy}.] 

This idea of baryomorphosis that the {\it observed}  baryon density is determined 
by a process similar to thermal relic WIMP freeze-out. The model does not explain the baryon asymmetry, but modifies a large initial baryon asymmetry into a final thermal WIMP-like baryon asymmetry; hence "baryomorphosis" rather than "baryogenesis".

\vs1
\noindent The ingredients of the baryomorphosis mechanism are: 

\vs1
\noindent $\bullet$ An initial baryon asymmetry, for example in a heavy particle $\Sigma$ which is out of thermal equilibrium, which decays at a low temperature $T_{d} \lae O(100) \GeV$. 

\vs1
\noindent $\bullet$ Pairs of new scalar particles,  $\pb$ and $\pbh$,  called  "annihilons", to which $\Sigma$ decays. The annihilons have opposite Standard Model (SM) charges but, crucially, do
\underline{not} have opposite baryon numbers. $\pb$ and $\pbh$ are distinct particles which can have different masses, although for simplicity we will consider their masses to be equal in the following.

\vs1
\noindent $\bullet$ A B-violating interaction which is broadly of weak interaction strength, via which $\pb$ and $\pbh$ annihilate, leaving a thermal WIMP-like baryon number density.

\vs1
\noindent $\bullet$  A mechanism to transfer the baryon asymmetry from annihilons to conventional quarks.

\vs1
 \noindent  This is illustrated schematically in Figure 1.

\begin{figure}[htbp]
\begin{center}
\includegraphics[width=170mm]{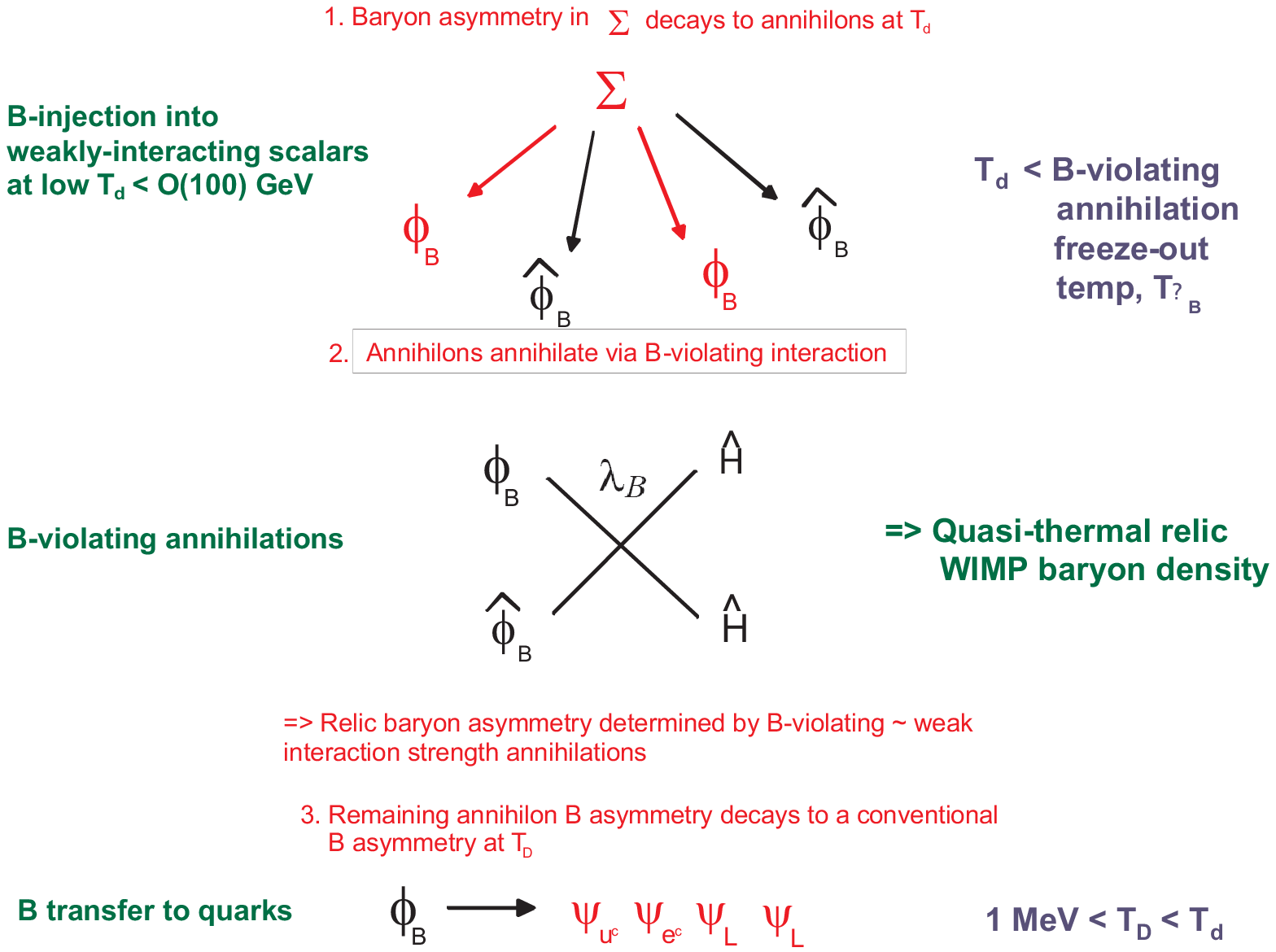}
\caption{}
\label{dsufig1}
\end{center}
\end{figure}

     In \cite{bm1} a simple model implementing the baryomorphosis mechanism was presented. In this model the annihilons annihilate to a complex boson $\hat{H}$ via a renormalizable B-violating interaction of the form 
\be{e1a} {\cal L}_{\phi_{B}\pbh \; ann} = \lambda_{B} \phi_{B} \pbh \hat{H}^{\dagger}\hat{H}  \;\;\; + \;h. \; c.  ~.\ee
Note that the $\hat{H}$ cannot be the Higgs boson, since the $\pb$, $\pbh$ will mix once the Higgs VEV is introduced, 
resulting in a B-violating mass insertion which will wash out the annihilon asymmetry via  scattering from thermal background quarks. Therefore the final state in the B-violating annihilation must have no VEV.

However, the original model in \cite{bm1} has some features which do not have a symmetry explanation and must therefore simply be imposed:

\vs1
\noindent  $\bullet$ There is no symmetry to prevent B-violating mass mixing terms like $\pb \pbh$ ($\Rightarrow$ B washout).

\vs1
\noindent $\bullet$ There is no suppression of renormalizable couplings of $\pb$ to SM fermions ($\Rightarrow$ $\pb$ decay too rapidly, before annihilating).

\vs1
\noindent In addition, although it is not really problem, no WIMP dark matter candidate is specified.

     In order to address these issues, in \cite{bm2} a model was presented which uses a simple discrete symmetry to ensure no $\pb$ $\pbh$ mixing terms leading to baryon washout. 
Since additional discrete symmetries are necessary to evade B washout, and since additional symmetries are also necessary to stabilize dark matter, it is natural to identify as dark matter particles the additional fields necessary as final states in the $\pb$ $\pbh$ annihilation process. The model can be naturally combined with gauge singlet (or inert doublet \cite{id}) dark matter to provide a unified model of baryomorphosis and scalar WIMP dark matter. 

      We introduce a pair of real singlet scalars $s$ $\sp$ plus a $Z_{2}$ discrete symmetry of the annihilons, $Z_{A}$,
\be{e3} {\rm Z_{A} :} \;\; \pb \rightarrow \pb  \; ; \;\;\; \pbh \rightarrow -\pbh   \; ; \;\;\;   s \rightarrow s  \; ; \;\;\;  \sp \rightarrow -\sp  ~,\ee 
with all SM fields invariant under $Z_{A}$.  This eliminates 
the dangerous B-violating mass mixing terms $\pb \pbh$ and $\pb \pbh H^{\dagger}H$ but allows the B-violating annihilation process
 \be{e4} {\cal L}_{\phi_{B}\pbh \; ann} = \lambda_{B} \phi_{B} \pbh s \sp     \;\;\; + \;h. \; c.  ~.\ee
Note that $\pb$ and $\pbh$ must carry gauge charges to prevent 
the dangerous terms $\pb \pb$ and $\pbh \pbh$, which are allowed under $Z_{A}$. 

We need a sceond discrete symmetry to stabilize the dark matter, $Z_{S}$,  
\be{e5} {\rm Z_{S} :} s \rightarrow - s  \; ; \;\;\;  \sp \rightarrow -\sp  ~.\ee 
The $Z_{A} \times Z_{S}$ symmetry then allows the coupling to the SM: 
\be{e6} \frac{\lambda_{s}}{2} s s H^{\dagger}H + \frac{\lambda_{\sp}}{2} \sp \sp H^{\dagger} H      ~.\ee 
This allows the $s$ and $\sp$ to annihilate down to a thermal WIMP-like dark matter density. 

  The formation of the final baryon asymmetry and dark matter density in this model is shown schematically in Figure 2. As before, the initial annihilon asymmetry is injected at a low temperature 
$T_{d} \sim 0.1-100 \GeV$. The $s$ and $\sp$ subsequently annihilate down to thermal WIMP-like density if $T_{d} < T_{s}$, or to a standard thermal WIMP density if $T_{d} > T_{s}$, where $T_{s} \approx m_{s}/25$ is the freeze-out temperature of the $s$ and $\sp$ dark matter.   

\begin{figure}[htbp]
\begin{center}
\includegraphics[width=150mm]{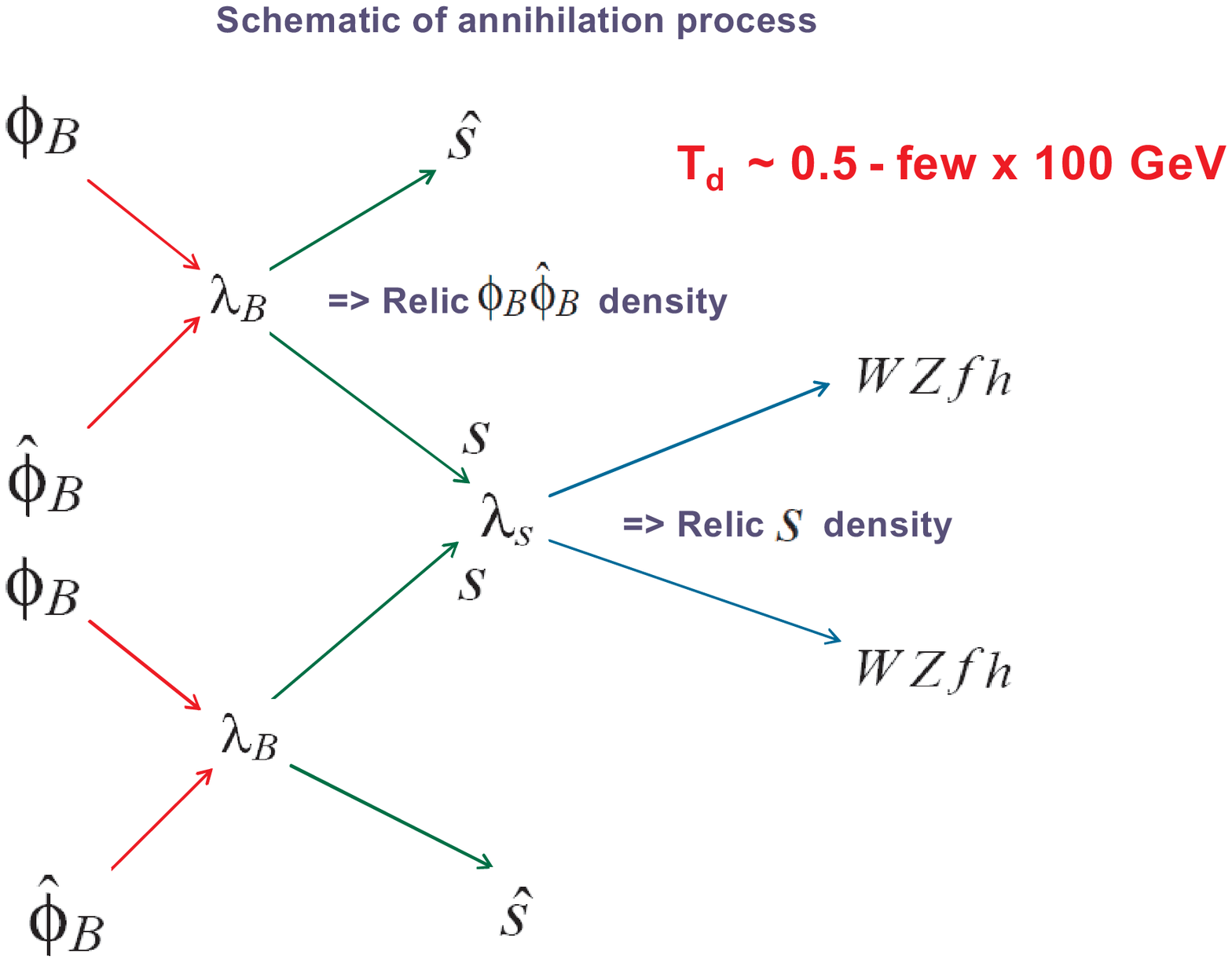}
\caption{}
\label{dsufig1}
\end{center}
\end{figure}

         The annihilation cross-section for $\pb \; \pbh \rightarrow s\;\sp$ is 
 \be{e5}    <\sigma v>_{\pb} = \frac{\lambda_{B}^2}{32 \pi m_{\pb}^2}
 \left( 1- \frac{m_{s}^{2}}{m_{\pb}^{2}} \right)^{1/2}    ~.\ee
The freeze-out number density at $T_{d}$ is then 
\be{e6}  n_{\pb}(T_{d}) \approx \frac{H(T_{d})}{<\sigma v>_{\pb}}    ~.\ee
(The $\pbh$ number density is the same when $m_{\pb} = m_{\pbh}$.) 
If $\pb$ later decays to baryon number $B(\pb)$ and $\pbh$ to $B(\pbh)$, the baryon asymmetry to dark matter ratio at present, $r_{BDM} \equiv \Omega_{B}/\Omega_{DM}$, is given by 
\be{e6a}  r_{BDM} = 3 (B(\pb) + B(\pbh)) \frac{m_{n}}{\Omega_{DM}} \frac{g(T_{\gamma})}{g(T_{d})^{1/2}} \left( \frac{4 \pi^3}{45 M_{Pl}^2} \right)^{1/2} \frac{T_{\gamma}^{3}}{\rho_{c}}  \frac{1}{T_{d}} \frac{1}{\left< \sigma v\right>_{\pb}}   ~.\ee
Here $g(T)$ is the effective number of relativistic degrees of freedom, $m_{n}$ is the nucleon mass, $\rho_{c}$ the critical density, $T_{\gamma}$ is the present photon temperature and $M_{Pl} = 1.22 \times 10^{19} \GeV$. The prefactor 3 accounts for the three colours of $\pb$. Requiring that $\Omega_{DM} = 0.23$ then determines the annihilon mass
\be{e6b} m_{\pb} = 2.81 \TeV  \times g(T_{d})^{1/4} r_{BDM}^{1/2}(B(\pb) + B(\pbh))^{-1/2}  \left(\frac{T_{d}}{1 \GeV}\right)^{1/2} \lambda_{B} \left(1 - 
\frac{m_{S}^{2}}{m_{\pb}^{2}} \right)^{1/4}    ~.\ee

   In this we have assumed that $\pb$ and $\pbh$ are non-relativistic when the 
B-violating annihilation occurs. This is generally true for gauge-charged $\pb$ and $\pbh$, as discussed in \cite{bm2}. It is also true in $s$ and $\sp$ annihilation provided $T_{d} \gae 0.4 \GeV$ \cite{bm2}.

    In Figure 3 we show the annihilon mass as a function of the B injection temperature $T_{d}$ for different values of the baryon-to-dark matter ratio
$r_{BDM}$. We have set $\lambda_{B} = \lambda_{s} = 0.1$, which are dimensionally natural values in particle physics models. Since the mystery of the baryon-to-dark matter ratio may be considered why they are within an order of magnitude of each other, we have shown $r_{BDM}$ in the range 0.1 to 10. We see that a wide range of annihilon masses in the 100 GeV to 10 TeV range is compatible with $r_{BDM}$ in the range
0.1 to 10 when $T_{d}$ is in the range 0.1 GeV to 100 GeV. 
There is also an upper bound on $T_{d}$ from the requirement that $T_{d} < T_{\pb}$, where $T_{\pb}$ is the freeze-out temperature of the B-violating annihilation process, $T_{\pb} \approx m_{\pb}/20$. $T_{\pb}$ equals a few hundred GeV for the range of $m_{\pb}$ considered in Figure 3.  

Therefore we see that for a wide range of $T_{d}$, $O(0.1) \GeV$  to  $O(100) \GeV$, the baryon density is naturally within an order of magnitude of the dark matter density. (Note that a low $T_{d}$ is essential for a baryon asymmetry to exist at all.) 

   In Figure 3 we have also shown lines for $m_{\pb}$ equal to 2 TeV and 3 TeV, corresponding to bounds that may be achieveable for coloured scalars at the LHC. The observed $r_{BDM} = 0.2$ favours annihilon masses less than $\sim $ 3 TeV over the range of $T_{d}$ from 0.1 to 100 GeV. Therefore, should the annihilons be coloured, there is a good prospect of producing them at the LHC. 

\begin{figure}[htbp]
\begin{center}
\includegraphics[width=150mm]{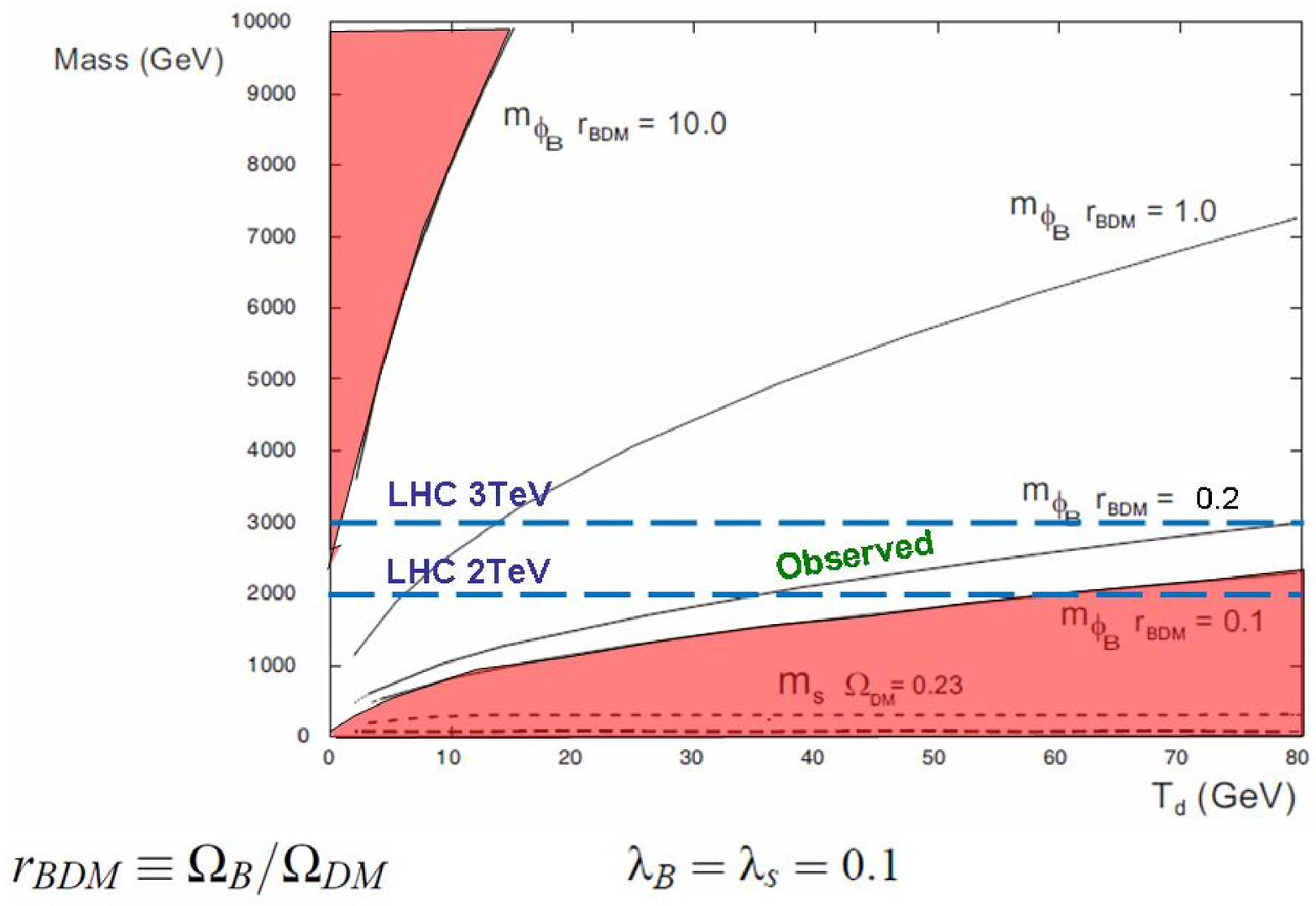}
\caption{}
\label{dsufig1}
\end{center}
\end{figure}

  In the case where $s$ and $\sp$ are degenerate in mass, both scalars are stable and contribute to dark matter. A key requirement is that $m_{s} < m_{\pb}$, so that the $\pb$ $\pbh$ annihilation process is kinematically allowed. If $T_{d} < T_{s}$, then the $s$ and $\sp$ dark matter density is non-thermal, coming from the $s$ and $\sp$ produced as final state particles in $\pb$ $\pbh$ annihilation. This gives an enhancement of the density by $T_{s}/T_{d}$ relative to the thermal relic $s$ density, 
\be{e15}  \Omega_{DM} =  \frac{2 m_{s}}{\rho_{c}} \frac{g(T_{\gamma})}{g(T_{d})^{1/2}} \left( \frac{4 \pi^3}{45 M_{Pl}^2} \right)^{1/2} \frac{T_{\gamma}^{3}}{T_{d}} \frac{1}{\left< \sigma v \right>_{s}}   ~.\ee
If $T_{d} > T_{s}$ then the dark matter is purely thermal relic in nature. In this case we can approximately replace $T_{d}$ by $T_{s}$ in \eq{e15}. 
The resulting $m_{s}$ is shown in Figure 4 for Higgs mass $m_{h} = 150 \GeV$, $\lambda_{S} = 0.1$ and $\Omega_{DM} = 0.23$. In Figure 5 we show $m_{s}$ together with $m_{\pb}$. For all $T_{d}$ the $s$ mass is much less than the $\pb$ mass, as required for consistency of the model. In addition, for most $T_{d}$ the $s$ freeze-out temperature is less than $T_{d}$, so that $s$ dark matter is in fact thermal relic in nature. Non-thermal $s$ dark matter is found only at $T_{d} \lae 10 \GeV$.

\begin{figure}[htbp]
\begin{center}
\includegraphics[width=150mm]{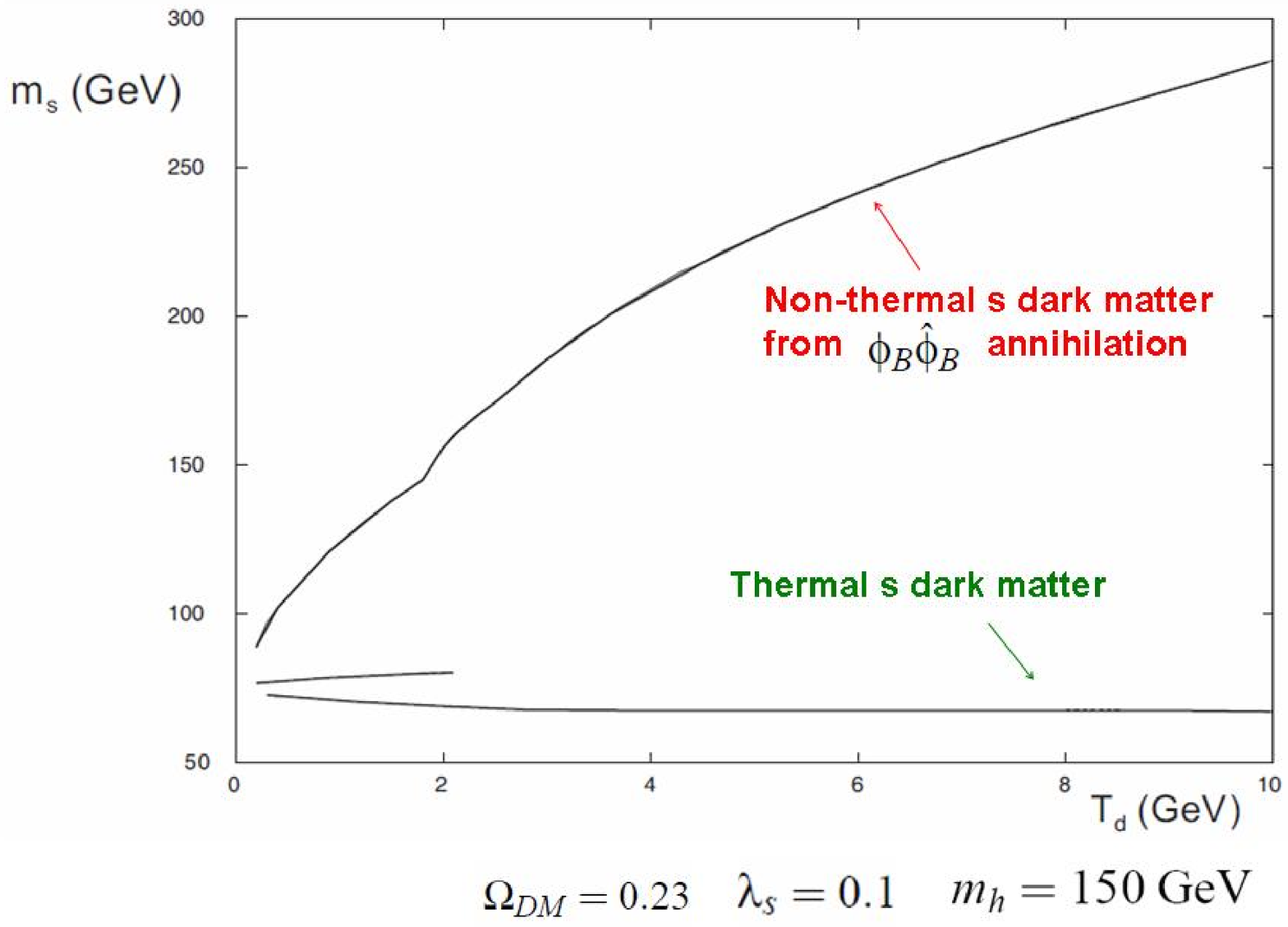}
\caption{}
\label{dsufig1}
\end{center}
\end{figure}

\begin{figure}[htbp]
\begin{center}
\includegraphics[width=150mm]{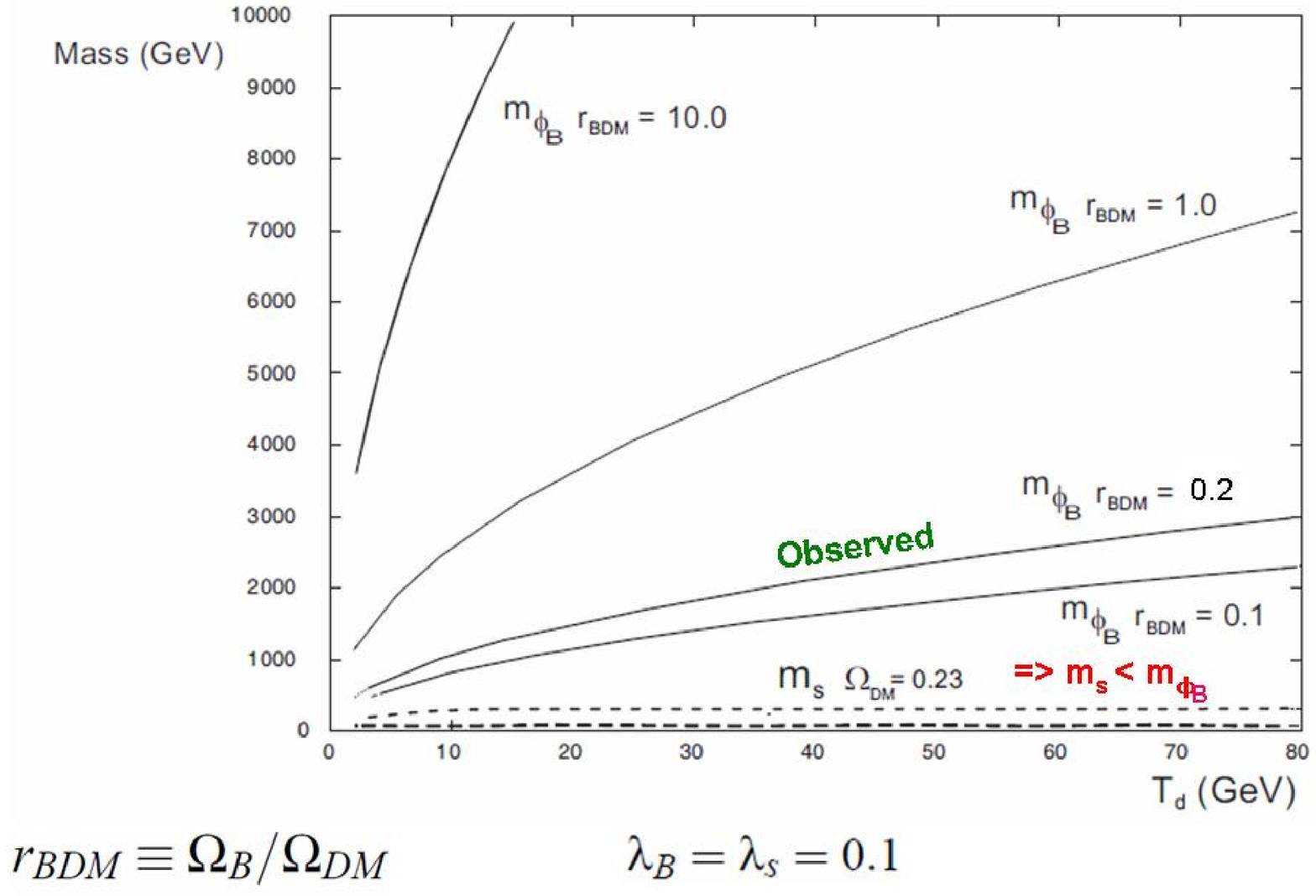}
\caption{}
\label{dsufig1}
\end{center}
\end{figure}

\vs1
    We conclude that a consistent baryomorphosis model can be constructed using only a 
simple $Z_{2} \times Z_{2}$ symmetry. 

\vs1
      So far we have produced a thermal WIMP-like baryon asymmetry in $\pb$ and $\pbh$ particles. This must still be transferred to conventional quarks. We will focus on the example of coloured annihilons, which have the best prospect of being produced at the LHC. We assume the annihilons transform under $SU(3)_{c} \times SU(2)_{L}$ as $\pb({\bf 3, 1})$ and $\pbh({\bf \overline{3}, 1})$. In order to produce a thermal WIMP-like baryon asymmetry, the decay of $\pb$, $\pbh$ must occur after the baryon asymmetry has frozen out, at $T_{D} < T_{d}$. 
This means that $\pb$ and $\pbh$ must be {\it very long-lived} particles: 
\be{e16}    1.5\; {\rm s} \gae \tau \gae 8 \times 10^{-11} \left(
\frac{100 \GeV}{T_{d}} \right)^2  \;\; s  ~,\ee
where the upper bound is from nucleosynthesis.
This provides a key experimental signature of annihilons. 
The long annihilon life-time requires either an extremely small renormalizable Yukawa coupling $\lambda \pb \overline{\psi} \psi$ to SM fermions $\psi$ with 
\be{e17} \lambda \lae 1.2 \times 10^{-10} \left( \frac{T_{d}}{1 \GeV}\right)
 \left( \frac{1 \TeV}{m_{\pb}}\right)^{1/2}   ~,\ee
  or a non-renormalizable coupling suppressed by a sufficiently large mass scale. The former possibility seems overtly unnatural, so we will consider the latter. However, in this case we need to explain why there are no renormalizable couplings leading to rapid $\pb$ decay.  

   The simplest way to achieve this is if $\pb$ has a large hypercharge e.g. $Y(\pb) = 5/3$. In this case $\pb$ and $\pbh$ can decay only via mass-suppressed non-renormalizable interactions. For example, for the case $\pb({\bf 3, 1}, 5/3)$ and $\pbh({\bf \overline{3}, 1}, -5/3)$, we can form
\be{e18} \frac{1}{M^{3}}\pb \overline{d_{R}^{c}} d_{R} \overline{L_{L}^{c}} L_{L}   ~\ee
and 
\be{e19}  \frac{1}{M^{3}}\pbh \overline{d_{R}} e_{R}^{c} Q_{L} Q_{L}  ~.\ee 
The mass $M$ should then be in the range $10^{6}-10^{8} \GeV$ to account for the low decay temperature $T_{D}$ \cite{bm1}. 

     Note that for $\pbh$ to decay, we must assume that $Z_{A}$ is slightly broken by the non-renormalizable operators. However, since these operators are suppressed by a large mass scale, this small breaking of $Z_{A}$ will not introduce any dangerous mass mixing between $\pb$ and $\pbh$.
If baryon number is conserved, then from \eq{e18} and \eq{e19} the baryon numbers\footnote{$\pb$ and $\pbh$ also carry lepton number, $L(\pb) = -2$ and $L(\pbh) = 1$.} of $\pb$ and $\pbh$ would be $B(\pb) = -2/3$ and $B(\pbh) = -1/3$.  

    However, if we do not assume baryon number conservation, then there are other possible operators, for example 
\be{e20} \frac{1}{M^{3}}\pb \left( \overline{e_{R}} Q_{L} \overline{e_{R}} L_{L} \right)^{\dagger}  ~,\ee
which allows $\pb$ to decay to a final state with $B = 1/3$. In this case the effective baryon number of $\pb$ and $\pbh$ will be determined by their dominant decay modes.

\vs1
\noindent Observation of such unstable particles might be achieved via the decay of particles which are stopped in detectors (see e.g. \cite{stopped}).

     In summary, the key experimental feature of the baryomorphosis model are the 
annihilons $\pb$, $\pbh$: 

\vs1
\noindent $\bullet$ Their mass must be 100 GeV - few TeV to produce weak-strength annihilations $\Rightarrow$ good prospect of producing at colliders. 

\vs1
\noindent $\bullet$ They have a long lifetime and decay to net B number. 

\vs1
\noindent $\bullet$ They may have B-violating decays.

\vs1
\noindent $\bullet$ Their long lifetime suggests large hypercharge $|Y| \geq 5/3$.

\vs1
\noindent $\bullet$ The best case is that of coloured annihilons: Can pair produce at LHC up to 2-3 TeV. (Similar to squark production.)

\section*{Conclusions}

    \vs1 
    \noindent $\bullet$ It is possible to {\it modify} a large initial baryon asymmetry to be similar to a thermal relic WIMP mass density $\Rightarrow$ {\it Baryomorphosis}. 

    \vs1
    \noindent $\bullet$ Requires additional particles and discrete symmetries 
$\Rightarrow$ Can naturally combine with a WIMP dark matter candidate (gauge singlet scalars, inert doublets) $\Rightarrow$ Simultaneous generation of thermal WIMP-like baryon and dark matter densities. 

    \vs1 
    \noindent $\bullet$ Therefore can understand \underline{both} of the puzzles of the baryon density: why it is similar to the dark matter density \underline{and} why it is similar to the "WIMP miracle" density. 

\vs1 
    \noindent $\bullet$ Need a low B-injection temperature, $T_{d} < {\rm few} \times 100 \GeV$.

\vs1 
   \noindent $\bullet$ Generically requires pairs of new particles ("annihilons") with gauge interactions and with mass $\sim 100 \GeV$ - few TeV $\Rightarrow$ Could be produced at the LHC.

\vs1 
\noindent $\bullet$ Two types of annihilon are necessary, with opposite gauge charge but \underline{different} B.

\vs1 
\noindent $\bullet$ Annihilons have long lifetime and decay to baryon number, with possibly large hypercharge and B-violating decay modes.   

\vs1  
\noindent To sum up, the WIMP miracle {\it can} account for DM without requiring either coincidence or anthropic selection to explain $\Omega_{B}/\Omega_{DM} = 1/5$.

\section*{Acknowledgement} 

This work is supported by the Lancaster-Manchester-Sheffield
Consortium for Fundamental Physics under STFC grant ST/J000418/1.


\end{document}